\documentclass[12pt]{iopart}
\usepackage{iopams}
\usepackage{epsfig}
\newcommand{\pTassoc}{\ensuremath{p_T^{assoc}}}
\newcommand{\pTtrig}{\ensuremath{p_T^{trig}}}
\newcommand{\pTab}{\ensuremath{<p_T^{assoc}<}}
\newcommand{\pTtb}{\ensuremath{<p_T^{trig}<}}

\begin{document}
\bibliographystyle{unsrt}
\title[Azimuthal Di-hadron Distributions from STAR]{Low- and Intermediate-p$_T$ Di-hadron Distributions in Au+Au Collisions at $\sqrt{s_{NN}}=200$ GeV from STAR}

\author{M. J. Horner for the STAR Collaboration}

\address{Lawrence Berkeley National Laboratory, One Cyclotron Road, Berkeley, CA 94720\footnote{Current address: UCT-CERN Research Centre, University of Cape Town, Rondebosch, 7701, South Africa}}
\ead{mjhorner@lbl.gov}
\begin{abstract}
We present a study of low- and intermediate-p$_T$ correlated azimuthal angular distributions in Au+Au collisions at $\sqrt{s_{NN}}=200\ \rm{GeV}$ from STAR. The near-side associated yields in Au+Au collisions are found to be strongly enhanced, due to contributions from large $\Delta\eta$. The enhancement is reduced for high \pTtrig. We show a strong broadening and enhancement of the away-side yield. The evolution of the away-side shape may be explained as the sum of a
broad structure from bulk response and a narrow peak from jet
fragmentation.
\end{abstract}


Azimuthal di-hadron correlation studies in d+Au and Au+Au collisions have
shown that hard partons from back-to-back scatterings interact strongly with the matter
that is generated and can be used to probe the medium~\cite{Adler:2002tq}. Different physics regimes can be probed by varying the $p_T$-range of the trigger and associated particles, \pTtrig\ and \pTassoc\ respectively. Previous work has shown that the away-side 
jet's interaction with the medium results in increased particle production at lower $p_T$~\cite{Adams:2005ph}, that a novel structure~\cite{Adler:2005ee,Jason} appears for specific combinations of \pTassoc\ and \pTtrig\ and that a clear di-jet signal can also been observed~\cite{Adams:2006yt} at the highest \pTtrig\ and \pTassoc.

The objective of this study is to use the large Au+Au data-set from the STAR experiment to study the interplay between different processes through a systematic study of the evolution of the correlated yields and shapes.

Di-hadron distributions are constructed using primary charged particles measured in the pseudo-rapidity range -1.0$<\eta<$1.0. Data are corrected for single particle efficiency and acceptance as well as the pair acceptance as a function of $\Delta\phi$. The uncorrelated background, which is modulated by elliptic flow ($v_2$), is removed by normalising the distribution $P(1+2\langle v_2^{trig}\rangle\langle v_2^{assoc.}\rangle\cos{(2\Delta\phi)})$ in the region $0.8<|\Delta\phi|<1.2$ and then subtracting it from the raw distributions. 

The nominal $v_2$ value subtracted is the mean of two measurements~\cite{Adams:2004bi,Adler:2002pu} using the reaction
plane and four-particle cumulant techniques which have different sensitivity to non-flow effects~\cite{Adams:2004bi}. The difference between the two $v_2$ results is used as the estimate of the systematic uncertainty in $v_2$.

The data presented in this paper are from collisions at $\sqrt{s_{NN}}=200\ \rm{GeV}$. 
Data were taken with increased luminosity over previous runs resulting in a significantly larger event sample allowing more differential studies than before. A total of 20M 0-12\% central Au+Au events were used. 
Reference results are shown from minimum bias d+Au collisions.  

The minimum bias trigger for d+Au collisions was defined by requiring that at least one beam-rapidity neutron impinge on the ZDC  in the Au beam direction. The measured minimum bias cross-section amounts to $95\pm3\%$ of the total d+Au geometric cross section~\cite{Adams:2003im}. Details of STAR triggering and reconstruction have been published previously~\cite{Ackermann:2002ad}.

\begin{figure}[hbt!]
\begin{minipage}[b]{0.5\linewidth} 
\includegraphics[width=.95\columnwidth]{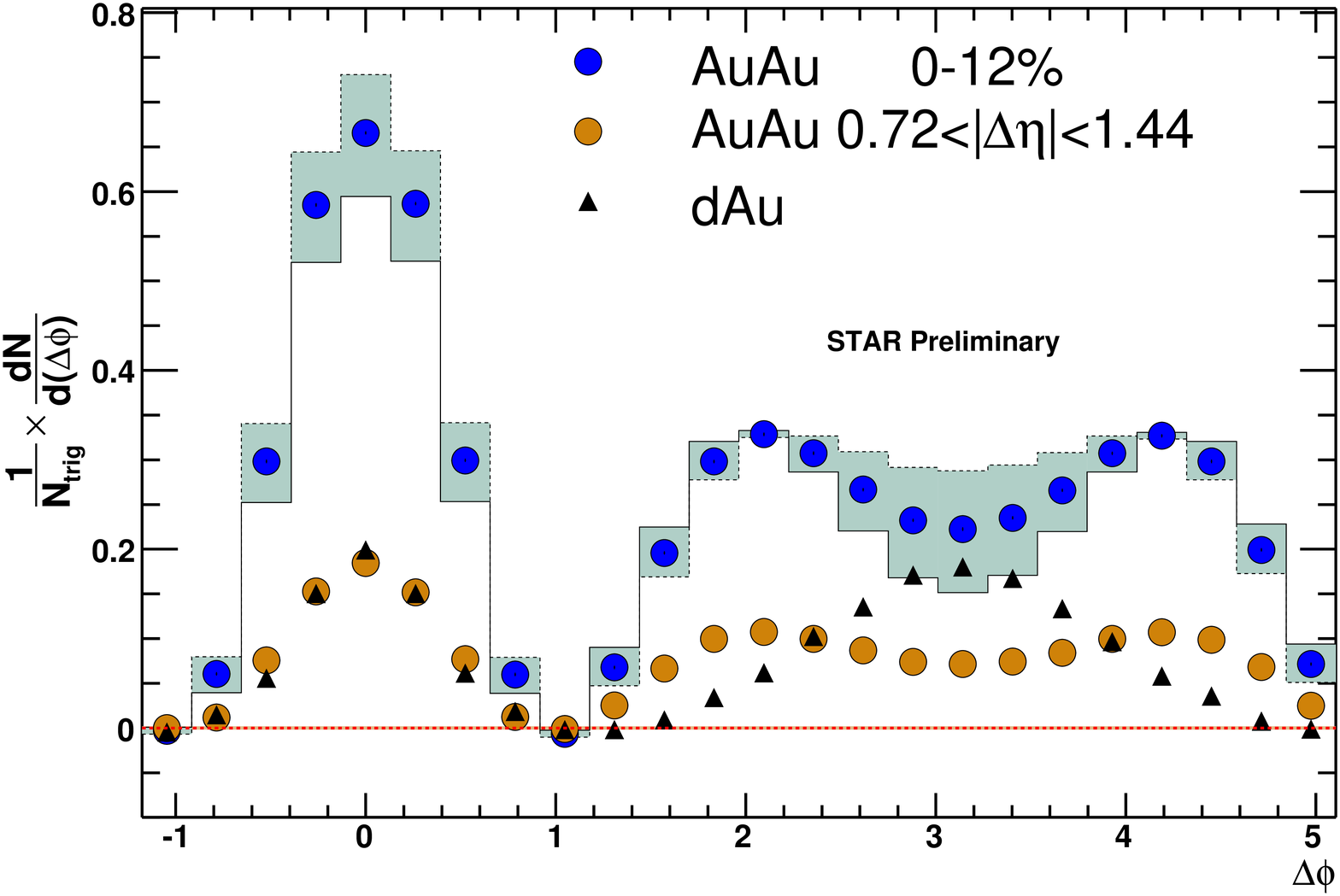}
\end{minipage}
\begin{minipage}[b]{0.5\linewidth} 
\caption{Azimuthal distributions for 1.0\pTab2.5~GeV/c and 2.5\pTtb4.0~GeV/c, for the full acceptance (blue circles) and for $0.72<|\Delta\eta|<1.44$ (orange circles) relative to the trigger particle. d+Au results (black triangles) are shown for reference. The bands around the data points show the systematic uncertainty from $v_2$ determination.}\label{fig:example}
\end{minipage}
\end{figure}

In figure~\ref{fig:example} we present the per-trigger di-hadron distributions for 1.0\pTab2.5~GeV/c and 2.5\pTtb4.0~GeV/c for the full acceptance of the TPC (blue circles) and for a case where the associated particle is restricted to a range in pseudo-rapidity of $0.72<|\Delta\eta|<1.44$ (orange circles) relative to the trigger particle. Full acceptance d+Au results (triangles) are included for reference. On the near-side there is a significant enhancement in yield relative to the d+Au reference. A significant fraction of this yield comes from a contribution at
$\Delta\eta > 0.7$. See \cite{Joern} for a more detailed study.

On the away-side there is a significant broadening of the shape and increase in the yield relative to d+Au. This difference in shape extends over a significant range in pseudo-rapidity as demonstrated by the large $\Delta\eta$ distribution. It is this modification of the away-side at intermediate $p_T$s which is studied here.

\begin{figure}[htb!]
\includegraphics[width=.95\columnwidth]{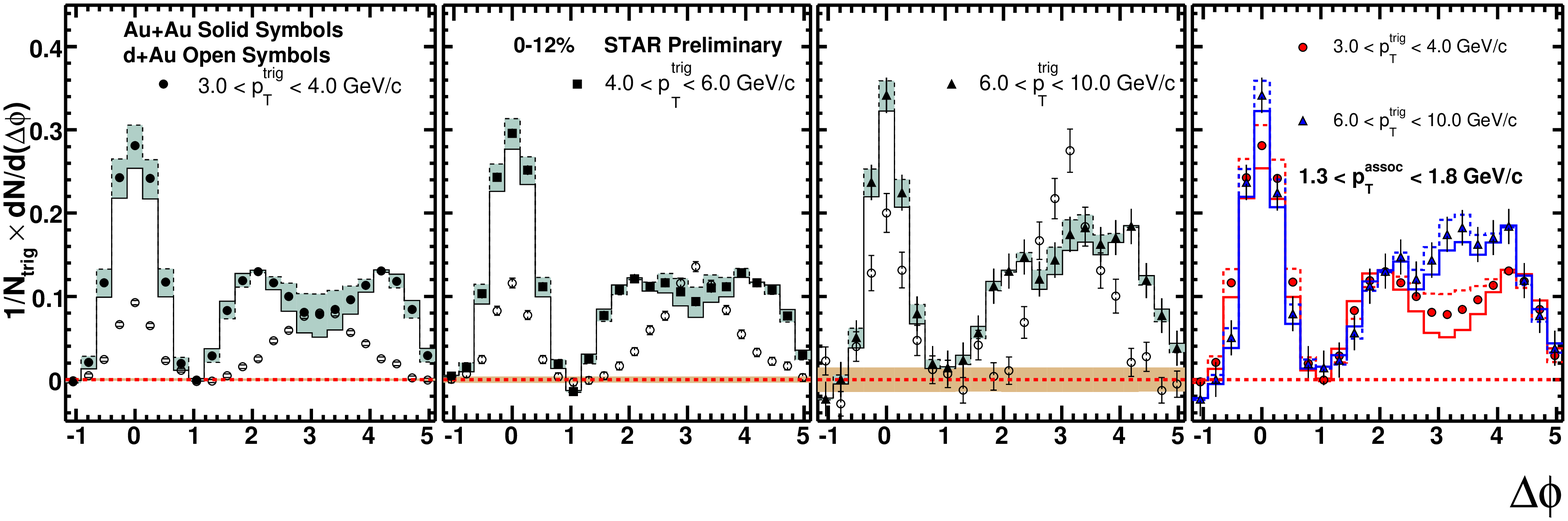}
\caption{Azimuthal distributions for 1.3\pTab1.8~GeV/c for 3 different choices of \pTtrig, 3.0\pTtb4.0~GeV/c (solid circles), 4.0\pTtb6.0~GeV/c (solid squares) and 6.0\pTtb10.0~GeV/c (solid triangles). d+Au results (open circles) are shown in the panels for reference.  The bands around the data points show the systematic uncertainty from $v_2$ determination while the band around zero shows the systematic uncertainty in the background pedestal determination. The right panel shows the superposition of the Au+Au results for 3.0\pTtb4.0~GeV/c and 6.0\pTtb10.0~GeV/c.}\label{fig:awaywidth}
\end{figure}

In figure~\ref{fig:awaywidth} we present per-trigger di-hadron distributions for a fixed \pTassoc\ range of 1.3 \pTab 1.8 GeV/c for different choices of \pTtrig. In d+Au collisions, the yield on the near-side increases with \pTtrig\ as expected from fragmentation. 

With increasing \pTtrig\ the away-side shape evolves to a flatter structure. The total width of the away-side structure is approximately independent of \pTtrig, as can be seen from the right-most panel. This could mean that there are two contributions to the signal: a broad structure with little or no dependence on \pTtrig, most likely from the medium response to the jet (as suggested in~\cite{Renk:2005si}), and a narrower jet-contribution that increases with \pTtrig.


\begin{figure}[htb!]
\begin{minipage}[b]{0.5\linewidth} 
(a)
\includegraphics[width=\linewidth]{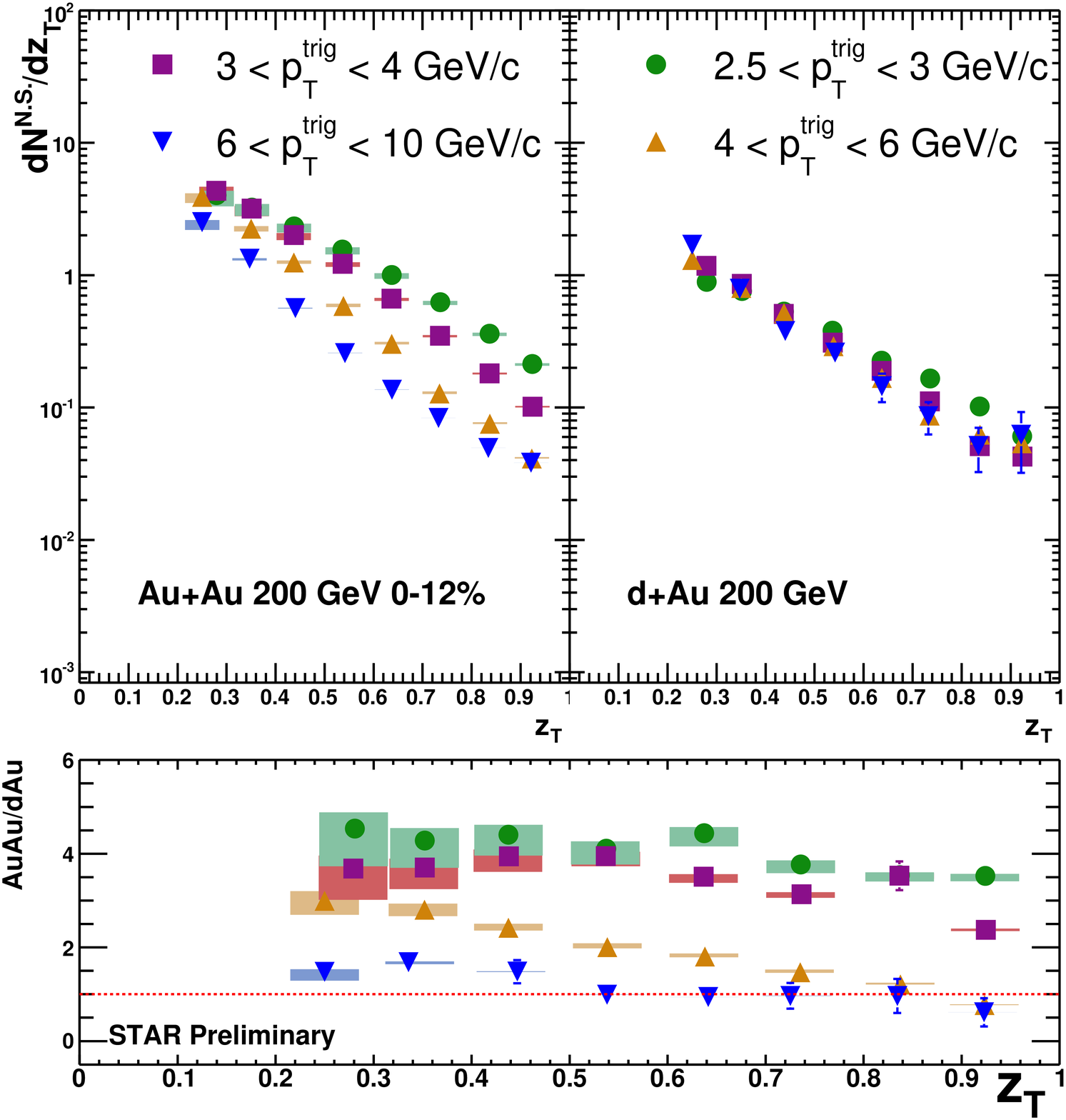}
\end{minipage}
\hspace{0.5cm} 
\begin{minipage}[b]{0.5\linewidth}
(b)
\includegraphics[width=\linewidth]{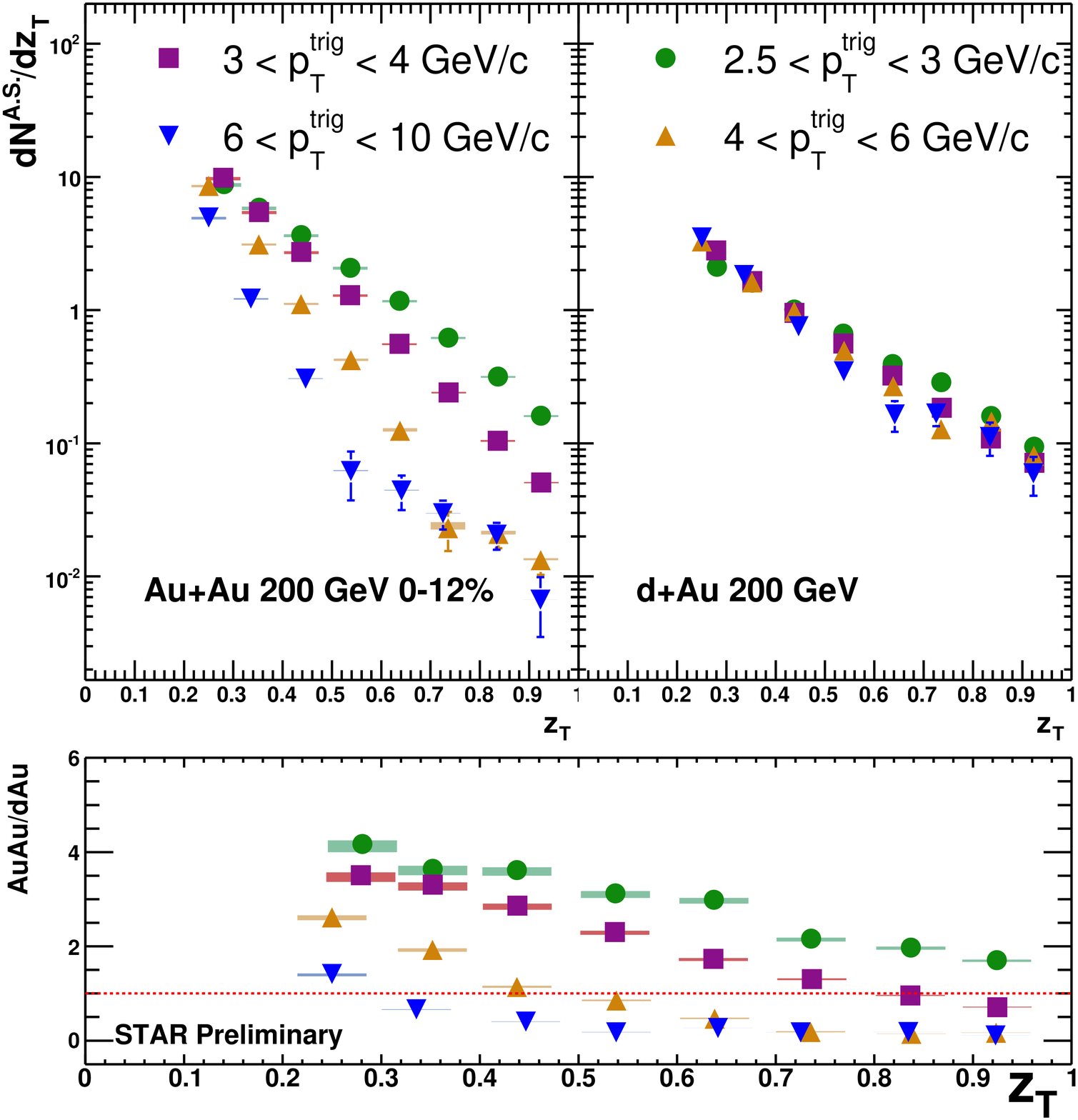}
\end{minipage}
\caption{Near- ($|\Delta\phi|<0.9$) and away-side yields ($|\Delta\phi|>0.9$), in figure (a) and (b) respectively, as a function of $z_T$ for Au+Au (left panel) and d+Au (right panel) for 2.5\pTtb3.0~GeV/c (circles), 3.0\pTtb4.0~GeV/c (squares), 4.0\pTtb6.0~GeV/c (triangles) and 6.0\pTtb10.0~GeV/c (inverted triangles). The bottom panel shows the ratio of Au+Au to d+Au results. The bands show the systematic uncertainty from $v_2$.}\label{fig:nsyields}
\end{figure}

Figure~\ref{fig:nsyields} summarises the dependence of the near- and away-side yields, ($|\Delta\phi|<0.9$) and ($|\Delta\phi|>0.9$) respectively, in d+Au and Au+Au collisions. The yields are given as a function of  $z_T$, where $z_T = \pTassoc/\pTtrig$, for different \pTtrig. The reference d+Au results are presented in the right panel. The near-side distributions, shown in (a), are steeper for increasing \pTtrig, with small changes in the case of d+Au. In the case of Au+Au the ordering is the same but with much larger differences for the different \pTtrig\ in the slope and total yield than for the d+Au results. The bottom panel shows the ratio of the Au+Au to d+Au results. The lowest \pTtrig\ data differ the most from the d+Au reference. As the \pTtrig\ increases the near-side approaches the d+Au results as is expected due to the increasing contribution from fragmentation.


In figure~\ref{fig:nsyields} (b) we summarise the away-side yields measured as a function of $z_T$ for different \pTtrig. The away-side yields for d+Au (right panel) are also very similar for all choices of \pTtrig. For the Au+Au results there is an inverse hierarchy with significantly lower yields for higher \pTtrig in the $z_T$ ranges shown. For high \pTtrig\ the data converge at high $z_T$.  This is more clearly seen in the ratio of Au+Au to d+Au results (bottom panel). Both sets of ratios reach ~0.2 and then remain constant. This happens at different $z_T$ for the two \pTtrig\ cases, but for a similar $p_T$ ($\sim$3 GeV/c). For lower $z_T$ values the yield increases steeply while for the lower two \pTtrig\ selections the yields are always much larger than 0.2 and have a significantly different slope and shape. 




In summary, we observe a large enhancement of the yields associated with trigger particles in Au+Au collisions, compared to a d+Au reference on both the near and away-side, indicating strong modifications to jet fragmentation in the hot and dense medium. The related enhancement is largest for the lowest \pTtrig and decreases for harder \pTtrig. On the near-side, the increase is partly located at large $\Delta\eta$ ($\Delta\eta>0.7$) and the yields approach the measurement in d+Au at the highest \pTtrig. On the away-side, a strong broadening is seen at lower \pTassoc, for all \pTtrig. The enhancement with respect to vacuum fragmentation on the away-side also decreases with \pTtrig, turning into a suppression at high-$p_T$. There seems to be a limiting value of this suppression at about 0.2. These results are qualitatively consistent with the idea that energy loss leads to a reduction of \pTtrig\ for a given jet energy. 
Increased yields and the changes in correlation shapes would then be due to fragmentation products of the radiated energy. Alternatively, one could imagine that the passage of high-$p_T$ partons excites the medium, leading to additional yield and modified correlation shapes. Quantitative modelling of the different physics processes is likely needed to distinguish the possible scenarios.

\section*{References}
\bibliography{Bibliography}

\end{document}